# Minimizing Grid Interconnection Capacity Requirements for AI Data Centers: A Developer-Side Planning Framework with Onsite Resources and Workload Flexibility

Hassan Zahid Butt, Rida Fatima, and Xingpeng Li

*University of Houston, 4222 Martin Luther King Blvd, Houston, 77204, Texas, USA*

**Abstract**

Securing grid interconnection capacity has become a bottleneck for AI data center projects and can take longer than constructing the facilities themselves. This mismatch can delay deployment for years, making early interconnection planning essential. This paper develops ICP-AI, an interconnection capacity planning framework from a data center developer's perspective. The framework minimizes grid import capacity under a prescribed onsite investment budget while jointly sizing photovoltaic (PV) and battery energy storage system (BESS) resources and scheduling deadline constrained workload flexibility. A secondary refinement fixes the minimum grid capacity and selects the minimum-investment PV-BESS portfolio among solutions that achieve that capacity. The framework is evaluated using monthly composite stress profiles across varying temporal assumptions, load shapes, flexible load fractions, and deferral windows. Results show that interconnection capacity reduction depends strongly on the planning environment: at a $100M budget, it is about 6% for the high load factor baseline, exceeds 10% under monthly average solar availability, and reaches 13.3% for a more diurnal load. At a $10M budget, 5% flexible load with a 1 h workload deferral window reduces BESS capacity from 15.30 to 4.87 MWh while increasing capacity reduction from 4.43% to 4.84%. To test sensitivity to temporal compression, the model is also solved over the full 8,760 h chronology, which preserves the main capacity and flexibility trends. Overall, ICP-AI quantifies the interconnection capacity and infrastructure substitution value of workload flexibility, providing an investment-interconnection frontier to support capital allocation and early project planning in constrained grid environments.

*Keywords:* AI data centers, battery energy storage systems, grid interconnection, load flexibility, mixed-integer linear programming, workload shifting.

## Nomenclature

| Symbol | Description |
|---|---|
| $m, d, h$ | Month, day, and hourly period indices |
| $a, k$ | Flexible load arrival hour and deferral index |
| $\Delta t$ | Time step duration (h) |
| $L_{m,h}$ | Native aggregate facility load (MW) |
| $CF_{m,h}$ | PV capacity factor (p.u.) |
| $c_{PV}$ | Capital cost factor for PV ($/MW) |
| $c_B$ | Capital cost factor for BESS ($/MWh) |
| $B$ | Onsite investment budget ($) |
| $\alpha$ | Flexible fraction of aggregate facility demand |
| $D$ | Maximum workload deferral window (h) |
| $M$ | Big-M constant for BESS operating state constraints |
| $S^{PV}$ | Installed PV capacity (MW) |
| $S^{B}$ | Installed BESS energy capacity (MWh) |
| $P^{IC}$ | Grid interconnection capacity variable; maximum permitted grid import (MW) |
| $P^{IC*}$ | Optimal $P^{IC}$ value from capacity optimization objective (MW) |
| $P_{m,h}^{grid}$ | Grid import power (MW) |
| $P_{m,h}^{curt}$ | Curtailed PV power (MW) |
| $P_{m,h}^{ch}$ , $P_{m,h}^{dis}$ | BESS charging and discharging power (MW) |
| $u_{m,h}^{ch}$ , $u_{m,h}^{dis}$ | BESS charging and discharging status variables |
| $H_B$ | BESS rated duration (h) |
| $E_{m,h}$ | Stored BESS energy (MWh) |
| $X_{m,a,k}$ | Flexible load arriving at hour $a$ and served $k$ hours later (MW) |
| $\eta_{RT}$ | BESS roundtrip efficiency |
| $\eta_{ch}, \eta_{dis}$ | BESS charging and discharging efficiencies |
| $C^{inv}$ | Installed PV-BESS capital investment cost ($) |
| $P_{m,h}^{nf}$ | Nondeferrable facility load (MW) |
| $P_{m,a}^{flex,arr}$ | Flexible load arrival power (MW) |
| $P_{m,h}^{flex,served}$ | Flexible load served in hour $h$ (MW) |
| $P_{m,h}^{served}$ | Total served facility load (MW) |
| $P^{fac,max}$ | Maximum facility power capacity (MW) |
| $d_m^{L}$ | Selected monthly peak load day |
| $d_m^{PV}$ | Selected monthly minimum PV energy day |
| $\phi(a, k)$ | Destination-hour mapping for a workload deferred by $k$ hours |

## 1. Introduction

Large AI data center projects are increasingly constrained by the time required to secure grid interconnection capacity rather than by facility construction alone. The International Energy Agency reports that data centers can be developed in roughly 1-3 years, whereas planning, permitting, and completing new grid infrastructure can require 5-15 years [1], [2]. Grid connection queues have reached record levels, and around 20% of global data center capacity planned through 2030 could face connection delays if current grid constraints are not addressed [1]. The urgency of this issue is also reflected in recent U.S. regulatory actions aimed specifically at accelerating the integration of data centers and other large loads [3], [4].

For a data center developer, this timing mismatch makes the required grid interconnection capacity a strategic project variable. A facility may have sufficient capital to deploy computing equipment and onsite energy resources yet still face a multi-year delay if its native peak demand triggers prolonged grid impact studies or requires substantial network reinforcement. Reducing the required interconnection capacity can thus improve the prospect of fitting a project within the capacity available at a candidate site and can reduce dependence on major grid upgrades. The relevant planning question is therefore not only how much energy a data center consumes, but how much grid capacity it requires and what onsite investment is needed to reduce that requirement.

AI workloads also create a flexibility opportunity that conventional large load planning does not fully exploit. Computing demand contains heterogeneous jobs with different latency and completion requirements, allowing a portion of batch oriented or deadline tolerant work to be shifted in time. Recent field demonstrations show that software-based orchestration can sustain power reductions without compromising service [5], [6]. Furthermore, production trace studies confirm that a meaningful fraction of data center demand is deferrable [7], [8]. This makes computational flexibility a potential planning resource rather than only an operational demand response mechanism.

Alongside this computational flexibility, onsite resources provide a complementary physical mechanism. Photovoltaic (PV) generation can reduce contemporaneous grid demand, while a battery energy storage system (BESS) can move energy from hours with available grid headroom to hours that would otherwise establish the interconnection peak. Prior work has studied workload shifting, renewable energy coordination, storage planning and dispatch, and market participation for data centers [9]-[20]. More recent studies have moved closer to the interconnection problem through flexibility aware siting, connect-and-manage operation, and AI data center infrastructure expansion [21]-[25].

Despite this progress, an important developer planning problem remains less developed. Operational studies commonly impose the interconnection limit as an input, whereas system level studies generally focus on network siting, expansion, or long-term demand growth. A data center developer instead needs a direct relationship between available onsite capital and the minimum grid interconnection capacity that can support a fixed facility load. The interaction between computational flexibility and BESS is equally important because both provide temporal shifting, but their marginal capacity value and their ability to substitute for one another need not evolve at the same rate.

This paper develops the Interconnection Capacity Planning for AI Data Centers (ICP-AI) framework to address this problem from a data center developer's perspective. ICP-AI minimizes the required grid import capacity while jointly sizing onsite PV and BESS under a prescribed capital budget. Aggregate workload flexibility is modeled by allowing eligible computation to be deferred within a prescribed time window without curtailing total work. A secondary minimum-investment refinement fixes this capacity optimum and identifies the lowest-investment PV-BESS portfolio that preserves it. The framework is evaluated using conservative monthly composite stress conditions, targeted sensitivity analyses, and a full 8,760 h chronological robustness check. Repeating the optimization across investment budgets quantifies the tradeoff between onsite investment and required interconnection capacity, while also revealing how workload flexibility reduces grid capacity requirements and substitutes for physical storage.

## 2. Literature Review

Data center flexibility has been studied from both computing and power system perspectives. Temporal and spatial workload shifting has been used to reduce electricity cost, provide demand response, support carbon free energy matching, and coordinate computing with renewable generation [8]-[15]. Temporal flexibility from periodic batch workloads has been quantified using production traces [8], while space-time load shifting flexibility has been formulated for electricity markets [9]. Additionally, spatiotemporal workload movement has been shown to improve hourly carbon-free energy matching [13]. More recent coordinated scheduling studies further examine how proactive workload transfers interact with the power system rather than treating computing and grid operation independently [14], [19], [20].

AI-specific studies strengthen the empirical basis for treating computational demand as flexible. A software-based control platform demonstrated on a 256 GPU cluster achieved a 25% power reduction for 3 h while maintaining quality of service requirements [5], and similar grid responsive computing capabilities have been validated on real world systems [6]. Furthermore, analysis of over one million tasks from Alibaba traces indicates that over 20% of estimated data center power is associated with tasks suitable for load shifting [7]. These studies support the use of an aggregate flexible load fraction and deferral window as planning parameters, provided they are interpreted as equivalent facility level flexibility rather than independent control of all electrical end uses.

Storage has also been investigated as a data center flexibility resource. Internet data centers and BESS have been jointly planned within a coupled smart grid framework [11], and storage sizing and dispatch have been optimized for grid flexibility services [12]. Emerging AI-focused studies increasingly coordinate BESS with computing demand by co-optimizing deadline constrained workloads and BESS dispatch under utility imposed peak limits [16], utilizing BESS as a buffering interface under connect-and-manage interconnection envelopes [17], and developing feedback optimization of storage to mitigate data center grid impacts [18]. These studies primarily address operation or compliance after a grid capacity limit or operating envelope has already been specified.

Data center interconnection planning is now becoming an active research area. Recent work develops a flexibility-aware, planner initiated siting framework explicitly motivated by multi-year interconnection queues [21]. Temporal and spatial AI workload flexibility has also been incorporated into generation and network capacity expansion planning to evaluate its effects on system investment, operating cost, and congestion [22]. Other frameworks formulate gigawatt-scale AI data center integration under connect-and-manage practices [23], while complementary work co-optimizes interconnection capacity, co-located renewable generation, and storage under uncertain long-term model scaling [24]. Policy-oriented work has also proposed explicit grid benefit requirements for connecting very large AI loads [25].

These studies establish the value of computational and electrical flexibility, but the project level decision considered here differs from both system capacity expansion and uncertainty-driven infrastructure planning. For instance, while Chen and Zheng examine how AI flexibility changes network wide expansion, system cost, and congestion [22], and Wang et al. examine how uncertain future training growth affects grid capacity decisions [24], ICP-AI instead considers a fixed proposed facility. It asks how much grid interconnection capacity can be reduced for a prescribed onsite investment budget while jointly optimizing PV, BESS, and deadline constrained computational flexibility. This developer perspective exposes the investment versus interconnection capacity tradeoff and allows the interaction and substitution between computational flexibility and physical storage to be quantified directly. Table 1 summarizes the distinction between ICP-AI and the most closely related data center interconnection planning studies.

Table 1. Comparison of ICP-AI with closely related data center planning studies

| Study | Perspective | Resources / flexibility | Primary planning question |
|---|---|---|---|
| [21] | System planner / siting | Flexible load service envelopes | Identify feasible data center sites under flexible grid service |
| [22] | System capacity expansion | Temporal and spatial AI workload flexibility | Quantify impacts on generation and network expansion |
| [24] | Stochastic expansion | Renewables, storage, and uncertain AI training load | Plan data center and grid capacity under load uncertainty |
| ICP-AI | Developer / fixed project | Joint PV-BESS sizing and deadline constrained workload flexibility | Minimize grid interconnection capacity for a prescribed onsite budget |

## *2.1 Research Gaps and Contributions*

The development of the ICP-AI framework is motivated by three primary research gaps. First, existing interconnection studies primarily take a system planner perspective or optimize broader infrastructure expansion, leaving the relationship between onsite developer investment and the minimum grid interconnection capacity required by a fixed data center project less explicit. Second, computational flexibility and BESS are commonly studied as operating resources, but their interaction and substitution within project level infrastructure planning are not systematically characterized. Third, the achievable interconnection capacity reduction can depend strongly on the temporal load and PV design conditions and on the native data center load shape, yet these sensitivities are rarely examined together from a developer planning perspective.

Against these gaps, the contributions of this paper are as follows.

1. This paper introduces ICP-AI, a developer-side planning framework that minimizes required grid interconnection capacity under a fixed onsite budget while jointly sizing PV and BESS resources.
2. Aggregate, deadline-constrained workload flexibility is integrated directly with infrastructure sizing, enabling the framework to quantify when computational load shifting complements or substitutes for battery storage.
3. Sensitivity analyses of temporal design conditions and native load shapes quantify how the achievable interconnection capacity reduction changes with the planning environment, distinguishing a conservative capacity planning basis from less restrictive temporal assumptions.
4. Targeted evaluations of flexible load fractions and deferral windows characterize diminishing returns to additional workload flexibility, explicitly distinguishing its interconnection capacity value from its infrastructure-substitution value.
5. A full 8,760 h chronological formulation is used to test the robustness of the principal capacity and substitution trends to temporal compression.

## 3. ICP-AI FRAMEWORK

ICP-AI represents a single grid connected AI data center campus with onsite PV and BESS, as illustrated in Fig. 1. Facility load and PV design profiles, technology assumptions, workload flexibility parameters, and an onsite investment budget are provided as planning inputs. The primary optimization determines the minimum grid import capacity while jointly sizing onsite resources and scheduling flexible workload. A secondary minimum-investment refinement then fixes that capacity and identifies the lowest-investment PV-BESS portfolio that preserves it. The final solution also determines hourly resource dispatch and workload allocation. Repeating the process across investment budgets forms an investment-interconnection frontier. Grid export and load shedding are not allowed, and all native and shifted demand must be served by grid imports, PV, and BESS.

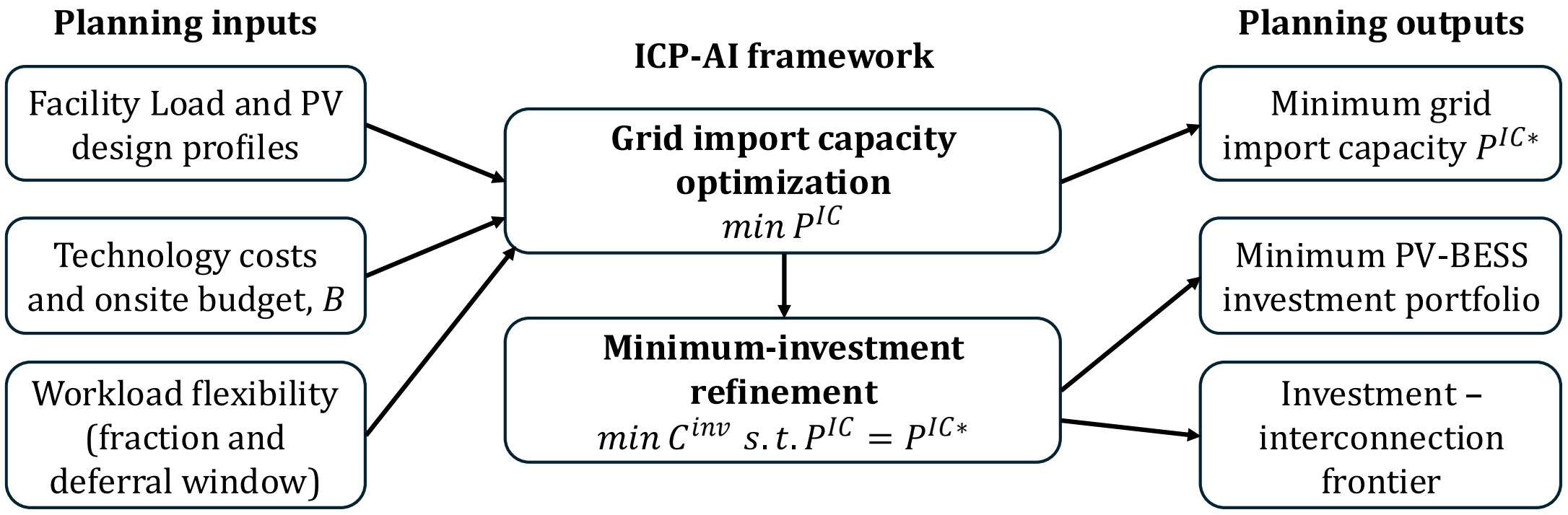

Fig. 1. ICP-AI framework and its principal planning inputs and outputs.

### *3.1 Grid Interconnection and Investment Formulation*

Let $S^{PV}$ and $S^{B}$ denote the installed PV capacity and BESS energy capacity, respectively. The principal planning variable, $P^{IC}$, represents the maximum grid import allowed at the facility point of interconnection; the terms grid interconnection capacity and grid import capacity are used interchangeably for this quantity. For each onsite investment budget $B$, the grid import capacity optimization solves:

$$\min\ P^{IC} \tag{1}$$

subject to the onsite capital investment constraint

$$c_{PV}S^{PV} + c_B S^{B} \leq B \tag{2}$$

and the hourly interconnection limit

$$0 \leq P_{m,h}^{grid} \leq P^{IC}, \quad \forall m, h \tag{3}$$

Let $P^{IC*}$ denote the minimum grid import capacity obtained above. The minimum-investment refinement then solves

$$\min\ C^{inv} = c_{PV}S^{PV} + c_B S^{B} \tag{4}$$

subject to

$$P^{IC} = P^{IC*} \tag{5}$$

All technical and operational constraints are identical in both problems. Because investment appears in the capacity optimization only as an upper bound, more than one PV-BESS portfolio may, in principle, attain the same minimum grid import capacity. The refinement therefore fixes $P^{IC} = P^{IC*}$ and selects the minimum-investment portfolio among those capacity-optimal solutions. If the refined portfolio uses less than the available budget, the unused budget cannot be used to obtain a lower $P^{IC}$ under the same modeled constraints; otherwise, $P^{IC*}$ would not be the optimal solution of the primary capacity minimization. For the cases evaluated in this study, the refinement process reproduces the capacity-optimal portfolios to numerical precision. It is retained to select the lowest-investment member of the capacity-optimal solution set if such multiplicity arises in other applications.

The resulting $P^{IC}$ should be interpreted as the minimum grid import capacity under the modeled design conditions, not as a universal connected load rating. Its interpretation from the developer perspective is most directly applicable where a utility or system operator recognizes an enforceable import cap, flexible service arrangement, or connect-and-manage structure supported by behind-the-meter resources and controllable demand. Large load interconnection rules remain jurisdiction specific and are currently evolving [3], [4], [17], [23].

The grid is not assigned an hourly energy price because ICP-AI is not an operating cost dispatch model. Instead, the grid can supply the required energy as long as hourly imports remain below $P^{IC}$, and the capacity optimization minimizes this limit. PV and BESS are selected only when their use reduces the maximum grid import under the available investment budget, while the secondary refinement removes any unnecessary investment without changing the minimum capacity requirement. The resulting dispatch should therefore be interpreted as a capacity feasible operating schedule rather than an economically optimized energy dispatch.

### *3.2 Aggregate Workload Flexibility*

Workload flexibility is represented at the aggregate facility level rather than by explicitly scheduling individual computing jobs. The native facility demand in month $m$ and hour $h$ is $L_{m,h}$. A fraction $\alpha$ is treated as equivalent flexible facility demand associated with rescheduling eligible information technology (IT) computation, while the remaining demand is nondeferrable:

$$P_{m,h}^{nf} = (1 - \alpha) L_{m,h} \tag{6}$$

$$P_{m,h}^{flex,arr} = \alpha L_{m,h} \tag{7}$$

This abstraction captures the facility level electrical effect of workload rescheduling and does not imply that cooling, lighting, or other auxiliary loads are independently controllable. Accordingly, $\alpha$ should be interpreted as an equivalent facility flexibility fraction rather than the fraction of every electrical end use that can be directly shifted.

For a workload arrival hour $a$, $X_{m,a,k} \geq 0$ denotes the portion of flexible demand assigned for service $k$ hours after arrival, where $k \in \{0, \dots, D\}$. Because $k$ is restricted to nonnegative values, workload can be deferred but not processed before its arrival and all flexible demand must be served within the prescribed deferral window:

$$\sum_{k=0}^{D} X_{m,a,k} = P_{m,a}^{flex,arr}, \quad \forall m, a \tag{8}$$

Equation (8) preserves the complete flexible workload, so temporal shifting does not represent load curtailment or unserved computation. For the representative day formulation, the destination hour is defined by $\phi(a,k) = 1 + mod(a - 1 + k,\ 24)$, with hours indexed from 1 to 24. This cyclic mapping ensures that workloads arriving near the end of a design day retain the same deferral window as workloads arriving earlier in the day. For example, a workload arriving at hour 23 with $D = 2$ may be served at hour 1 of the next repetition of the same design condition. Because $k \geq 0$, this treatment does not permit early processing; it only avoids artificially truncating the deferral window at midnight. The full 8,760 h formulation uses chronological workload allocation and does not require this repeated-day boundary treatment. The flexible demand served in hour $h$ is therefore

$$P_{m,h}^{flex,served} = \sum_{\substack{a,k: \\ \phi(a,k)=h}} X_{m,a,k} \tag{9}$$

Total facility demand served in each hour is

$$P_{m,h}^{served} = P_{m,h}^{nf} + P_{m,h}^{flex,served} \tag{10}$$

A facility power limit prevents workload shifting from creating an infeasible instantaneous facility demand:

$$P_{m,h}^{served} \leq P^{fac,max}, \quad \forall m, h \tag{11}$$

### *3.3 PV, BESS, and Power Balance*

PV generation is represented using the hourly capacity factor $CF_{m,h}$. Because grid export is not permitted, PV curtailment is allowed when available generation exceeds the amount that can be used by the facility or BESS:

$$0 \leq P_{m,h}^{curt} \leq CF_{m,h} S^{PV} \tag{12}$$

The hourly facility power balance is

$$P_{m,h}^{dis} + CF_{m,h} S^{PV} + P_{m,h}^{grid} = P_{m,h}^{served} + P_{m,h}^{ch} + P_{m,h}^{curt} \tag{13}$$

The BESS is represented as a fixed duration storage resource, with energy capacity $S^B$ as the planning variable. For a rated duration $H_B$, the corresponding charging and discharging power limits are $S_B/H_B$. Binary variables $u_{m,h}^{ch}$ and $u_{m,h}^{dis}$ prevent simultaneous charging and discharging.

$$u_{m,h}^{ch} + u_{m,h}^{dis} \leq 1 \tag{14}$$

$$P_{m,h}^{ch} \leq M u_{m,h}^{ch}, \qquad P_{m,h}^{ch} \leq \frac{S^B}{H_B} \tag{15}$$

$$P_{m,h}^{dis} \leq M u_{m,h}^{dis}, \qquad P_{m,h}^{dis} \leq \frac{S^B}{H_B} \tag{16}$$

Here, $M$ is a sufficiently large constant that activates the charging and discharging state constraints. The round-trip efficiency $\eta_{RT}$ is represented symmetrically through the charging and discharging efficiencies:

$$\eta_{ch} = \eta_{dis} = \sqrt{\eta_{RT}} \tag{17}$$

Stored energy evolves according to

$$E_{m,h} = E_{m,h-1} + \left(\eta_{ch} P_{m,h}^{ch} - \frac{P_{m,h}^{dis}}{\eta_{dis}}\right) \Delta t \tag{18}$$

$$0 \leq E_{m,h} \leq S_B, \forall m, h \tag{19}$$

The stored BESS energy is bounded by the installed energy capacity. Each monthly design day uses the same initial and terminal stored-energy level, which prevents free stored energy and artificial energy transfer between independently modeled monthly conditions. Grid charging of the BESS is permitted because interconnection capacity can be reduced by shifting grid energy from lower-demand periods to hours that would otherwise determine the maximum grid import requirement.

### *3.4 Monthly Composite Design Conditions*

The primary planning model uses twelve monthly 24 h design profiles, corresponding to 288 modeled hours. Annual weighting is not applied because the objective is to determine a grid capacity requirement under selected stress conditions rather than to estimate annual energy use or operating cost. For each month, the load profile is selected as the day containing the highest hourly facility demand in the annual load series:

$$d_m^L = \underset{d \in m}{argmax} \; \underset{h}{max} L_{d,h} \tag{20}$$

The PV profile is selected independently as the day with the minimum integrated daily PV availability within the same month:

$$d_m^{PV} = \underset{d \in m}{argmin} \sum_{h=1}^{24} CF_{d,h}^{PV} \Delta t \tag{21}$$

The 24 h load profile from $d_m^L$ is then paired with the 24 h PV profile from $d_m^{PV}$ to form one composite stress condition for each month. Because the load and PV days are selected independently, they need not correspond to the same calendar day. The resulting profile is therefore a deliberately conservative synthetic condition rather than a coincident operating day. Minimum daily PV availability is used rather than PV output at the single peak load hour because BESS can charge before the binding grid import period and transfer that energy through time. Daily PV availability therefore better represents the renewable energy available to support capacity reduction over the full design day. This composite construction is used as the conservative baseline design condition. Section 4 defines alternative temporal conditions that progressively relax the load and PV pairing assumptions, allowing the sensitivity of the resulting interconnection capacity requirement to be quantified.

# 4. CASE DESCRIPTION AND EXPERIMENTAL DESIGN

## *4.1 Data Center Load and PV Profiles*

The case represents a 100 MW nominal IT AI data center with a fixed power usage effectiveness (PUE) of 1.2. Annual IT load profiles are generated using the Lawrence Berkeley National Laboratory Data Center and Industrial Electricity Load Shape Maker [26]. The tool generates parametric annual profiles using several configurable characteristics, including data center usage type, diurnal pattern, and short-term variability. For data center applications, the predefined usage types include training, inference, and mixed workloads, while the diurnal patterns range from nearly flat demand to business-hour and customer-driven profiles. These dimensions are specified separately in the tool. In this study, the Mixed usage type is used for both load cases, while the Flat and Business designations describe their different diurnal patterns.

The primary Mixed Flat profile represents a highly utilized mixed-workload facility with limited time-of-day variation. After scaling to facility demand, its annual average load is 83.274 MW, its peak is 87.915 MW, and its load factor is 94.72%. A second Mixed Business profile is used to evaluate sensitivity to native load shape. It retains the same mixed usage type but introduces a stronger business-hour diurnal pattern. Its annual average demand is similar at 84.339 MW, while its substantially higher peak of 99.619 MW results in a lower load factor of 84.66%. This provides a controlled comparison between a nearly flat, high load factor profile and a more diurnal profile without changing the nominal IT capacity or PUE. The contrast between the two native load shapes is illustrated in Fig. 2.

Hourly PV capacity factors are obtained from a full year PVWatts profile for Houston, Texas [27]. The annual capacity factor series ranges from 0 to 0.90, with a mean hourly value of 0.1703. The baseline model applies the monthly composite design procedure described in Section 3.4 independently to all twelve months. Figure 2 shows the corresponding D1 load and PV profiles for January, April, July, and October. These four months are presented for clarity, while all twelve-monthly design conditions are retained in the optimization.

The study is framed around AI data centers because the application is motivated by large AI-oriented facilities and by the temporal flexibility of computational workloads considered in the formulation. The optimization itself, however, does not depend on a specific training or inference workload archetype. It can be applied to other data center types by supplying the corresponding facility load profile and workload-flexibility parameters. More generally, the same formulation can represent other large flexible loads when their demand profile, flexible fraction, and allowable shifting window can be characterized.

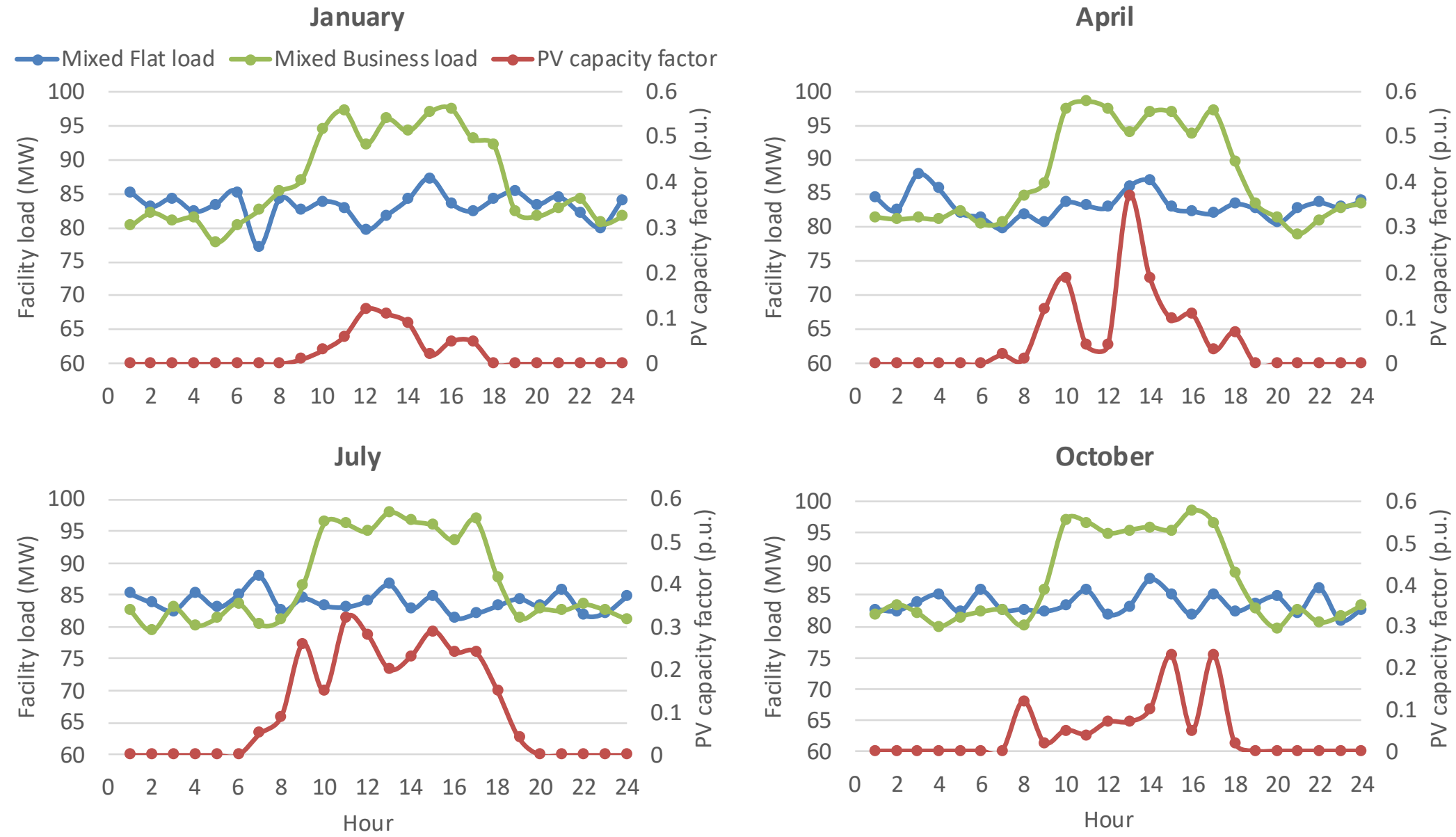

Fig. 2. Representative seasonal load and PV profiles for the Mixed Flat and Mixed Business cases constructed using the D1 temporal design procedure

### *4.2 Planning and Technology Assumptions*

The baseline planning and technology assumptions are summarized in Table 2. PV and BESS capital costs are based on the adopted 2026 cost projections from the National Renewable Energy Laboratory (NREL) Annual Technology Baseline [28].

Table 2. Planning and technology assumptions

| Parameter | Value |
|---|---|
| Nominal IT capacity | 100 MW |
| PUE / facility power limit | 1.2 / 120 MW |
| Baseline load | Mixed Flat, load factor = 94.72% |
| PV location | Houston, Texas |
| PV capital cost | $1.432M/MW |
| BESS capital cost | $0.47674M/MWh |
| BESS duration | 4 h |
| BESS round-trip efficiency | 0.90 |
| Baseline workload flexibility | 5% of aggregate facility load |
| Baseline deferral window | 1 h |
| Grid export / load shedding | Not allowed / not allowed |
| Grid charging of BESS | Allowed |
| Primary temporal resolution | 12 × 24 h monthly composite profiles |
| Robustness resolution | 8,760 chronological hours |

Electricity tariffs, operating revenues, and payments for workload flexibility are intentionally excluded because ICP-AI isolates the relationship between upfront onsite investment and the required interconnection capacity. The linear investment cost representation and fixed 4 h BESS duration are capacity planning assumptions rather than a lifecycle economic model; BESS degradation, augmentation, and operation and maintenance costs are outside the present scope.

### *4.3 Baseline Cases and Sensitivity Design*

Two baseline configurations are used throughout the study. C1 represents the case without workload flexibility, while C2 allows 5% of aggregate facility load to be deferred by up to 1 h. The 5% flexible load fraction and 1 h deferral window are selected as deliberately moderate assumptions relative to the field demonstrations and production trace studies reported in [5]-[8]. Table 3 summarizes the complete experimental design, detailing the baseline configurations, sensitivity scenarios, and chronological robustness check. The first sensitivity (S1) considers four temporal profile constructions (D1-D4) to evaluate the planning environment. D1 serves as the conservative peak load and minimum PV composite used as the primary planning basis; D2 uses the PV profile historically coincident with each monthly peak load day; D3 retains the monthly peak load profile but uses monthly hourly average PV availability; and D4 uses monthly hourly average load and PV profiles as a typical operating benchmark. The subsequent sensitivities vary one modeling dimension at a time while retaining the baseline assumptions. S2 tests the sensitivity of the framework to the native load shape by comparing the default Mixed Flat profile against a Mixed Business profile. Finally, S3 and S4 systematically vary the flexible load fraction ($\alpha$) and deferral window ($D$) to characterize diminishing returns to additional workload flexibility, deferral value saturation, and infrastructure substitution.

Table 3. Experimental design

| **Study** | **Configurations / values** | **Purpose** | **Budgets ($M)** |
|---|---|---|---|
| Baseline | C1: no flexibility; C2: 5% flexibility, $D = 1$ h | Investment-interconnection frontier | 0-100 |
| S1: Temporal condition | D1: conservative; D2: coincident PV; D3: average PV; D4: average load and PV | Temporal design sensitivity | 0, 10, 50, 100 |
| S2: Load shape | Mixed Flat vs Mixed Business; D1; C1/C2 | Native load shape sensitivity | 0, 10, 50, 100 |
| S3: Flexible load fraction | $\alpha = 0, 2.5, 5, 10, 15, 20\%$; $D = 1$ h | Flexibility level and storage substitution | 0, 10 |
| S4: Deferral window | $D = 0, 1, 2, 3, 6$ h; $\alpha = 5\%$ | Deferral value and saturation | 0, 10 |
| Robustness | Full 8,760 h chronology; C1/C2 | Sensitivity to temporal compression | 0, 10, 50, 100 |

### *4.4 Implementation and Verification*

The models are implemented in Pyomo and solved with Gurobi 12.0.3 as mixed-integer linear programs (MILPs) on an Intel Xeon W-2195 CPU at 2.30 GHz with 128 GB RAM. Each 288 h planning case first determines the minimum grid import capacity and then applies the minimum-investment refinement described in Section 3.1. A post-optimality check across representative configurations confirmed that the refinement preserves the capacity optimum and reproduces the corresponding infrastructure portfolios to numerical precision for the cases evaluated.

All reported cases are solved with a zero requested optimality gap and are checked for power balance, grid capacity, terminal SOC, facility capacity, and flexible workload allocation residuals. The 288 h design cases solve to optimality in under one minute on this hardware. The full 8,760 h formulation, evaluated separately as a chronological robustness check of the interconnection capacity trends, also solves to optimality within a minute.

## 5. RESULTS AND DISCUSSION

### *5.1 Baseline Investment and Interconnection Capacity Frontier*

Fig. 3 presents the relationship between onsite investment and minimum grid import capacity for C1 and C2, while Table 4 reports selected capacity reductions and the corresponding PV-BESS portfolios. All PV and BESS capacities reported in Sections 5.1-5.4 correspond to the minimum-investment portfolios associated with the optimized grid capacity solutions. The native Mixed Flat facility peak is 87.915 MW. With no onsite investment, allowing 5% of aggregate facility demand to shift by up to 1 h reduces the required grid capacity to 84.975 MW, corresponding to a 2.940 MW or 3.34% reduction without PV or BESS. This result isolates the direct interconnection capacity value of workload flexibility.

Onsite investment provides a steep initial reduction in grid capacity followed by strong diminishing returns. Without workload flexibility, a $10 million budget lowers the required capacity to 84.021 MW, capturing 3.894 MW of the 5.340 MW reduction achieved at $100 million, or 72.9% of the high-budget benefit. With workload flexibility, the total capacity reduction achieved at $10 million is 77.8% of that obtained at $100 million. These results show that most of the achievable capacity reduction is obtained in the lower part of the investment range, with progressively smaller gains at higher budgets.

Workload flexibility changes the infrastructure mix more strongly than it changes the final grid capacity. At a $10 million budget, C1 installs 1.889 MW of PV and 15.301 MWh of BESS. C2 instead installs 5.361 MW of PV and only 4.872 MWh of BESS, reducing required storage by 68.2% while lowering $P^{IC}$ by an additional 0.363 MW. Flexible computation performs part of the short duration temporal shifting that would otherwise require battery storage, allowing a larger share of the available investment to be allocated to PV.

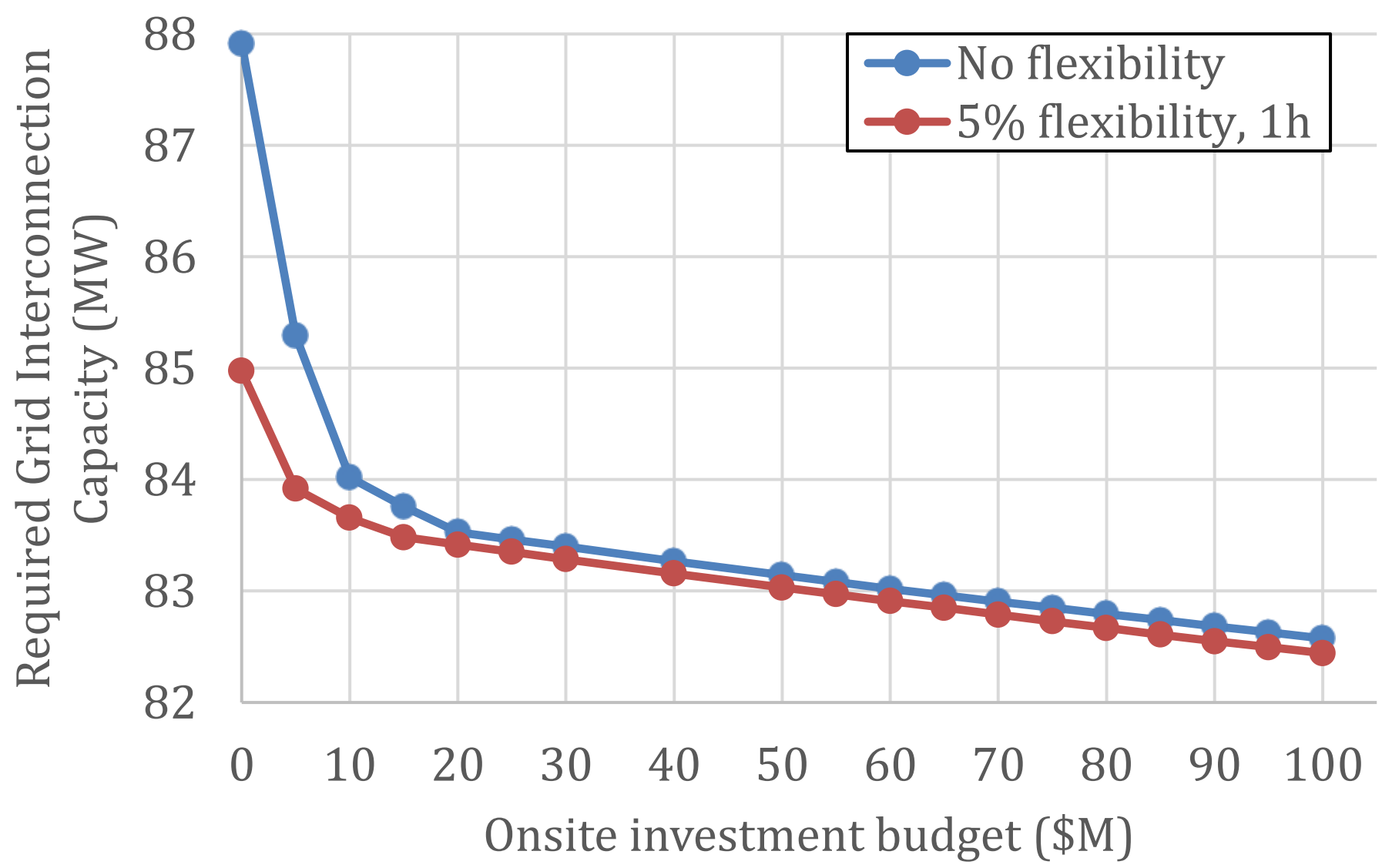

Fig. 3. Minimum required grid interconnection capacity as a function of onsite investment for the baseline C1 and C2 configurations.

At higher investment levels, the additional grid capacity benefit of workload flexibility becomes small because PV, BESS, and computational shifting increasingly overlap in the peak reduction service they provide. At $100 million, C1 and C2 reduce the grid capacity requirement by 6.07% and 6.22%, respectively. This does not mean that workload flexibility has lost its planning value. C2 requires only 18.99 MWh of BESS compared with 33.17 MWh in C1, showing that computational flexibility continues to substitute for physical storage even after its direct interconnection capacity benefit has largely saturated.

Table 4. Baseline ICP-AI results

| Budget ($M) | $P^{IC}$-C1 (MW) | Reduction C1 (%) | PV-C1 (MW) | BESS-C1 (MWh) | $P^{IC}$-C2 (MW) | Reduction C2 (%) | PV-C2 (MW) | BESS-C2 (MWh) |
|---|---|---|---|---|---|---|---|---|
| 0 | 87.915 | 0.00 | 0.00 | 0.00 | 84.975 | 3.34 | 0.00 | 0.00 |
| 10 | 84.021 | 4.43 | 1.89 | 15.30 | 83.658 | 4.84 | 5.36 | 4.87 |
| 50 | 83.143 | 5.43 | 28.44 | 19.47 | 83.032 | 5.55 | 31.13 | 11.39 |
| 100 | 82.575 | 6.07 | 58.79 | 33.17 | 82.443 | 6.22 | 63.51 | 18.99 |

### *5.2 Temporal Design Condition Sensitivity*

The D1 baseline is intentionally conservative because the monthly peak load profile is paired with the minimum PV availability profile even when the two occur on different calendar days. At a $100 million budget, D1 yields interconnection capacity reductions of 6.07% for C1 and 6.22% for C2. When the PV profile observed on the monthly peak load day is used instead, as in D2, the reductions increase to 7.85% and 8.09%, respectively. This corresponds to an additional 1.56–1.64 MW of achievable capacity reduction and shows that the assumed temporal pairing of load and PV materially affects the interconnection requirement. Fig. 4

compares the resulting capacity reductions across the four temporal design conditions at a $100 million budget.

D3 retains the same monthly peak load profiles as D1 and D2 but replaces the conservative PV profile with the monthly hourly average PV profile. This isolates the effect of solar availability while preserving the load stress condition. At $100 million, the resulting capacity reductions increase to 10.39% for C1 and 10.67% for C2. The result shows that the rapid saturation observed under D1 is not an inherent limitation of PV and BESS. Under low solar availability, the value of additional investment is increasingly constrained by limited renewable energy, whereas improved PV availability allows additional investment to continue reducing the grid capacity requirement.

The preferred infrastructure mix also changes substantially with the temporal design condition. At $100 million, C1 installs 58.79 MW of PV and 33.17 MWh of BESS under D1, 53.46 MW of PV and 49.17 MWh of BESS under D2, and 42.35 MW of PV and 82.54 MWh of BESS under D3. Greater solar availability progressively reduces the PV nameplate capacity needed to provide useful energy, allowing more of the available investment to be allocated to storage for temporal redistribution.

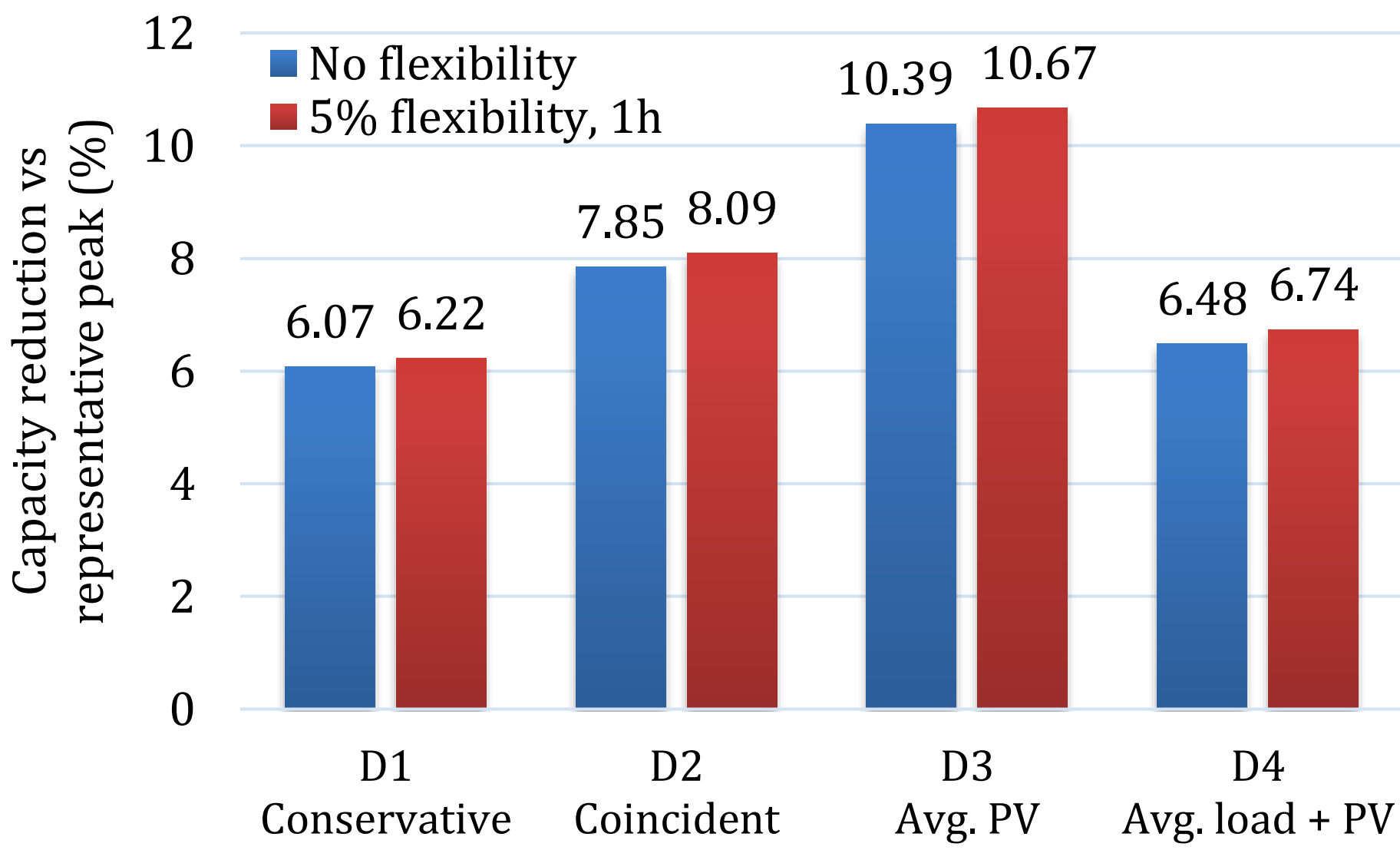

Fig. 4. Interconnection capacity reduction relative to each temporal condition's representative peak at a $100 million budget. D4 represents a typical operating benchmark rather than the conservative capacity planning basis.

D4 uses monthly hourly average load and PV profiles and therefore represents a typical operating benchmark rather than a conservative capacity planning condition. Its representative peak falls to 83.821 MW, compared with the 87.915 MW annual peak of the Mixed Flat profile. Accordingly, the D4 percentages in Fig. 4 are evaluated relative to its own representative peak and should not be interpreted as reductions from the actual annual peak. D4 produces reductions of 6.48% for C1 and 6.74% for C2 at $100 million, with a C1 portfolio of 42.78 MW of PV and 81.27 MWh of BESS. Hourly averaging suppresses short duration load peaks and makes an already flat profile even smoother, leaving less temporal variation for BESS and workload flexibility to exploit. Across all four temporal conditions, the difference between C1 and C2 remains modest at this high investment level; the dominant sensitivity in Fig. 4 is therefore the temporal load-PV design condition rather than the incremental effect of workload flexibility.

### *5.3 Load Shape and Load Factor Sensitivity*

Fig. 5 compares the interconnection capacity reduction obtained for the Mixed Flat and Mixed Business load profiles with and without workload flexibility. As illustrated earlier in Fig. 2, the Mixed Flat baseline has an annual load factor of 94.72%, with only a 4.64 MW difference between its average and peak demand. The Mixed Business profile has a similar annual average demand of 84.34 MW but a substantially higher peak of 99.62 MW, resulting in a lower load factor of 84.66% and a 15.28 MW peak-to-average gap. The larger peak-to-average spread visible in Fig. 2 creates more short-duration peak demand that can be shifted or supported by onsite resources.

At a $100 million budget, the Mixed Flat profile achieves capacity reductions of 6.07% for C1 and 6.22% for C2. Applying the same D1 temporal design procedure and investment budget to the Mixed Business profile yields reductions of 13.31% and 13.37%, respectively. The achievable relative reduction therefore more than doubles for the more diurnal load profile, showing that the value of onsite resources depends strongly on the temporal structure of the native facility demand.

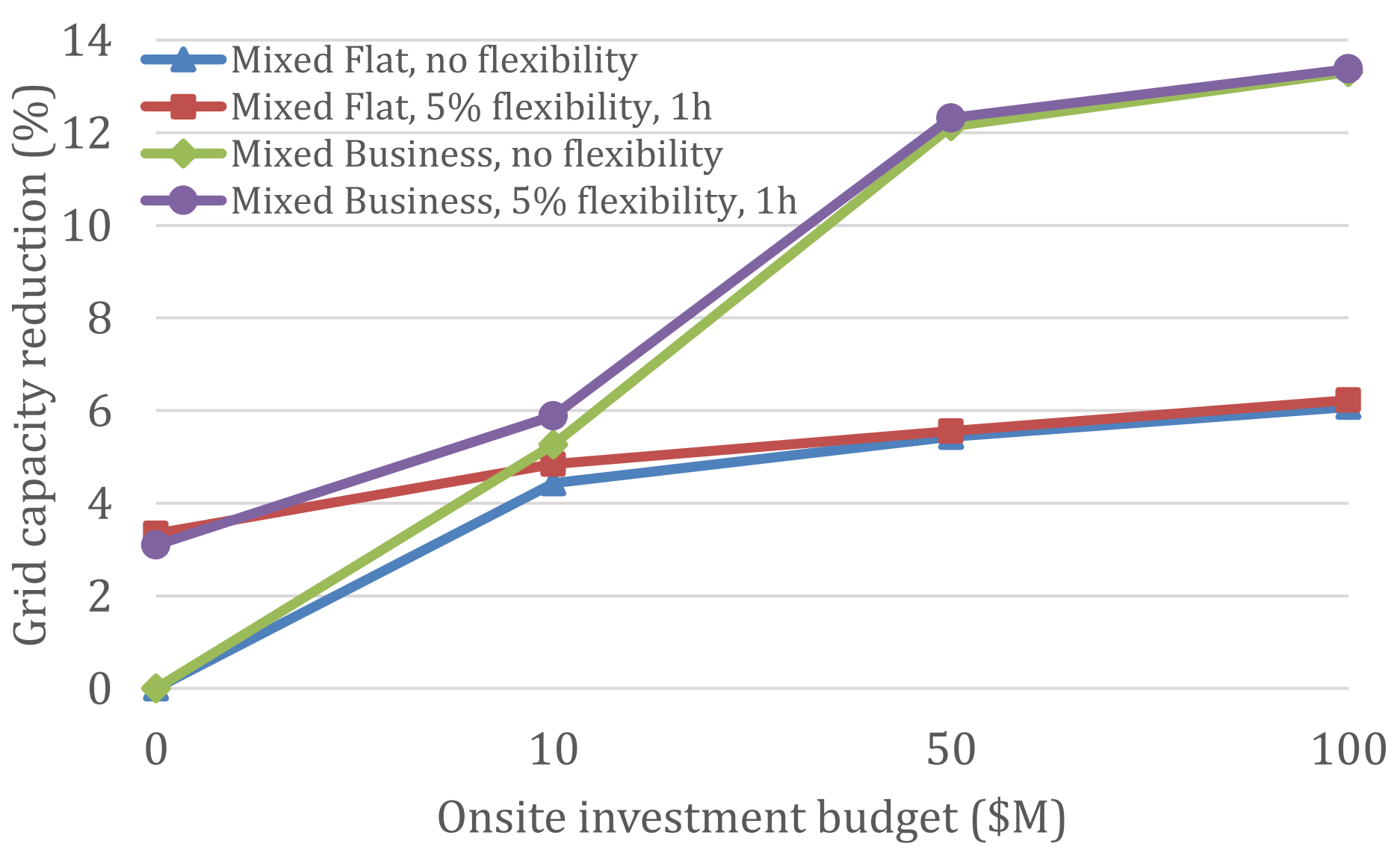

Fig. 5. Effect of native load shape on grid interconnection capacity reduction.

The $10 million Mixed Business case provides a clear physical interpretation of this effect. Under C1, the optimal portfolio contains no PV and 20.976 MWh of BESS. With the assumed 4 h duration, this corresponds to 5.244 MW of battery power, exactly matching the reduction from the 99.619 MW native peak to 94.375 MW. The initial investment is therefore allocated entirely to BESS-based peak reduction. Under C2, the same PV and BESS capacities are retained, while workload flexibility lowers the required grid capacity by a further 0.620 MW. At this investment level, computational flexibility therefore complements rather than substitutes for physical storage.

The interaction changes at higher investment levels. At $50 million, workload flexibility reduces BESS capacity from 101.09 to 90.76 MWh while increasing PV capacity from 1.26 to 4.70 MW. At $100 million, BESS capacity decreases from 67.50 to 58.70 MWh while PV increases from 47.36 to 50.29 MW. The nonmonotonic BESS trajectory reflects a transition from direct battery peak reduction at lower budgets to a combined PV and storage solution as additional storage alone becomes less effective at reducing the interconnection requirement.

### *5.4 Flexible Workload Sensitivity*

S3 varies the flexible load fraction from 0% to 20% while holding the maximum deferral window at 1 h. Fig. 6 shows the resulting interconnection capacity reduction at $0 and $10 million investment levels together with the corresponding optimized BESS capacity at $10 million. With no onsite investment, increasing flexibility reduces the required grid capacity from 87.915 MW at 0% flexibility to 84.097 MW at 15%, after which no further reduction is observed. The maximum flexibility-only benefit is therefore 3.818 MW, or 4.34% of the native facility peak.

Most of the direct capacity benefit is captured at relatively modest flexibility levels. The 5% baseline reduces the no-investment grid capacity requirement by 2.940 MW, corresponding to 77.0% of the maximum benefit observed at 15–20% flexibility. At a $10 million budget, the same 5% flexibility level captures 70.2% of the maximum incremental capacity benefit relative to the no-flexibility case.

The infrastructure substitution effect is considerably stronger than the remaining change in grid capacity. At a $10 million budget, BESS capacity decreases from 15.301 MWh with no flexibility to 8.081 MWh at 2.5%, 4.872 MWh at 5%, and only 0.413 MWh at 10% flexibility. At 15% flexibility, BESS is fully displaced, while PV capacity increases from 1.889 to 6.983 MW. Between 5% and 10% flexibility, the grid capacity requirement decreases by only 0.143 MW, yet BESS capacity falls by another 4.46 MWh. This shows that workload flexibility can continue to reshape the minimum investment infrastructure portfolio even after most of its direct interconnection capacity value has already been captured.

A complementary sensitivity S4 fixes the flexible load fraction at 5% and varies the maximum deferral window from 0 to 6 h. The numerical results are summarized in Table 5. With no onsite investment, extending the deferral window from 0 to 1 h provides a 2.940 MW reduction in required grid capacity, capturing 77.0% of the maximum observed scheduling benefit. A 2 h window captures 97.1%, and no further capacity reduction is obtained beyond 3 h. At a $10 million budget, the corresponding shares are 70.2% for a 1 h window and 97.9% for a 2 h window.

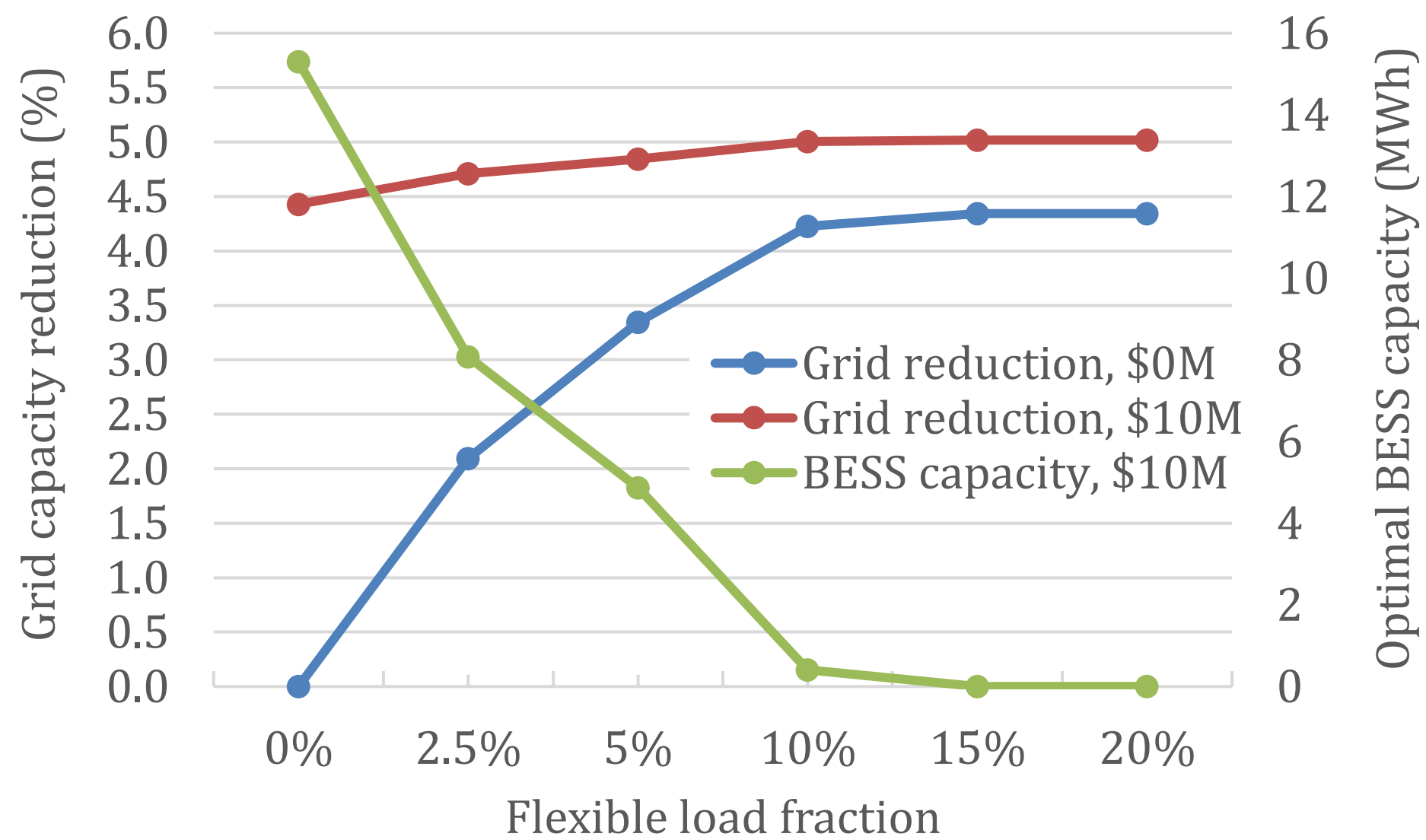

Fig. 6. Grid interconnection capacity reduction at $0 and $10 million budgets and optimized BESS capacity at $10 million as a function of flexible load fraction.

The effect on physical storage is even more pronounced. At a $10 million budget, increasing the deferral window from 0 to 1 h reduces BESS capacity from 15.301 to 4.872 MWh. Extending the window to 2 h leaves only 0.391 MWh of BESS, and at 3 h storage is fully displaced. Over the same 1–3 h range, however, the grid capacity requirement improves by only 0.154 MW. Thus, additional scheduling time provides progressively smaller direct interconnection capacity benefits while continuing to substitute strongly for battery storage.

Table 5. Workload flexibility sensitivity at a $10 million budget

| Study | Flexible load (%) | Deferral (h) | $P^{IC}$ (MW) | Reduction (%) | PV (MW) | BESS (MWh) |
|---|---|---|---|---|---|---|
| S3 | 0 | 1 | 84.021 | 4.43 | 1.889 | 15.301 |
| S3 | 2.5 | 1 | 83.774 | 4.71 | 4.293 | 8.081 |
| S3 | 5 | 1 | 83.658 | 4.84 | 5.361 | 4.872 |
| S3 | 10 | 1 | 83.515 | 5.01 | 6.846 | 0.413 |
| S3 | 15 | 1 | 83.503 | 5.02 | 6.983 | 0.000 |
| S3 | 20 | 1 | 83.503 | 5.02 | 6.983 | 0.000 |
| S4 | 5 | 0 | 84.021 | 4.43 | 1.889 | 15.301 |
| S4 | 5 | 1 | 83.658 | 4.84 | 5.361 | 4.872 |
| S4 | 5 | 2 | 83.514 | 5.01 | 6.853 | 0.391 |
| S4 | 5 | 3 | 83.503 | 5.02 | 6.983 | 0.000 |
| S4 | 5 | 6 | 83.503 | 5.02 | 6.983 | 0.000 |

Several configurations in Table 5 converge to identical solutions. Once the flexible fraction reaches 15% at a 1 h window (S3), or the deferral window reaches 3 h at 5% flexibility (S4), workload shifting alone is sufficient to remove the short-duration component of the binding grid import period. Beyond this point, BESS is fully displaced and the $10 million budget is fully allocated to PV (6.983 MW). The remaining 83.503 MW requirement is therefore determined by hours that cannot be further reduced through the permitted workload shifting under the fixed-budget solution.

Together, the S3 and S4 results show that the selected 5% flexible load fraction and 1 h deferral window lie in a moderate region of the sensitivity space. This baseline captures most of the available direct capacity benefit while retaining meaningful contributions from PV, BESS, and workload flexibility. Larger flexibility fractions or longer deferral windows produce progressively smaller additional grid-capacity reductions and increasingly substitute for short-duration battery storage.

### *5.5 Full Year Chronological Robustness*

To test whether the principal interconnection capacity findings are sensitive to the 288 h temporal compression, C1 and C2 are also evaluated using the complete 8,760 h load and PV chronology. Unlike the monthly composite formulation, the annual model carries BESS SOC continuously across the year and imposes a single year end terminal condition. Workload deferral is also modeled chronologically rather than cyclically within independent 24 h design conditions. Fig. 7 compares the resulting capacity reductions with those obtained from the compact monthly formulation.

The full year formulation produces somewhat larger achievable capacity reductions but preserves the main trends. At a $100 million budget, the C1 reduction increases from 6.07% under the conservative monthly formulation to 7.22% under full chronology, while C2 increases from 6.22% to 7.28%. At $10 million, the corresponding reductions are 4.43% versus 4.86% for C1 and 4.84% versus 5.16% for C2. Thus, the full chronology changes the magnitude of the reduction moderately but does not alter the conclusions regarding diminishing investment returns or the incremental capacity value of workload flexibility.

The larger difference appears in the resulting storage capacities. Under full chronology, interday SOC carryover allows the BESS to transfer energy across the actual annual sequence, whereas the monthly composite formulation requires each independent design day to return to its initial SOC. This additional temporal freedom allows larger storage capacities to be used more effectively and therefore affects the resulting infrastructure mix more strongly than the grid capacity requirement itself. Because the 8,760 h model is used primarily as a robustness check of the interconnection capacity trends, its PV-BESS mix is not compared with the minimum-investment portfolios obtained from the compact planning formulation.

The comparison therefore indicates that the headline interconnection conclusions are relatively robust to temporal compression, while detailed PV-BESS sizing is more sensitive to chronological representation.

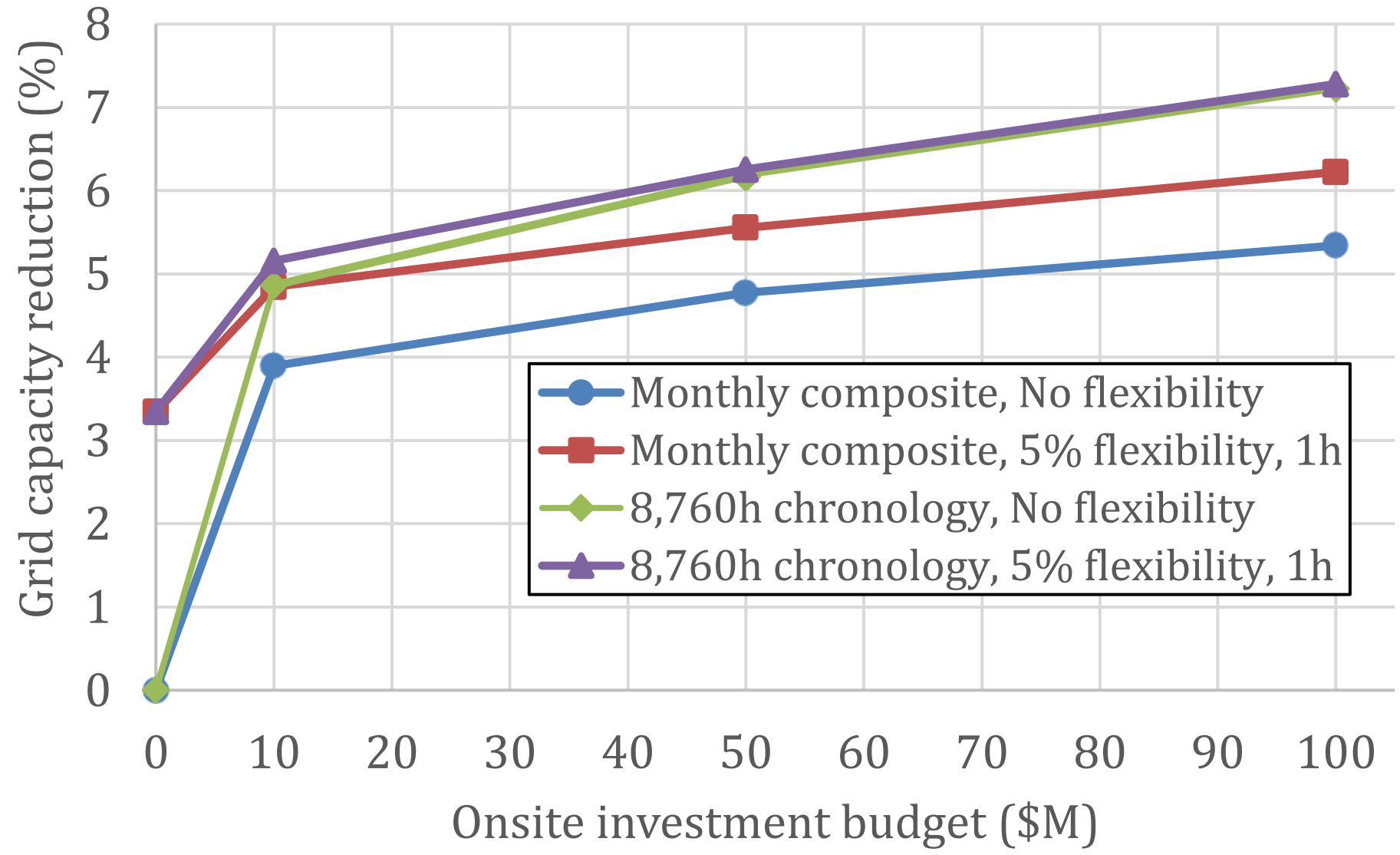

Fig. 7. Interconnection capacity reduction under the monthly composite and full 8,760 h chronological formulations.

### *5.6 Planning Implications*

The combined results show that the value of onsite resources for reducing data center interconnection requirements cannot be represented by a single capacity reduction percentage. Under the conservative baseline, the reduction at high investment is about 6%, while historically coincident solar conditions increase it toward 8%, monthly average solar availability raises it above 10%, and the more diurnal Mixed Business profile produces reductions above 13%. These differences show that the achievable interconnection benefit depends strongly on both the temporal design condition and the native facility load shape.

Workload flexibility should also be valued through both the grid capacity it can avoid and the physical infrastructure it can displace. At moderate flexibility levels, computational shifting can complement BESS, as observed in the $10 million Mixed Business case where the same battery portfolio supports a lower grid capacity requirement. As the flexible load fraction or deferral window increases, workload shifting increasingly substitutes for BESS and reallocates investment toward PV. Its direct interconnection capacity benefit can therefore become small even while its infrastructure substitution value remains substantial.

The investment-interconnection frontier can also support site screening from the developer perspective. If the grid capacity presently available at a candidate point of interconnection is known, the frontier can be read in reverse to identify the minimum modeled onsite investment needed to bring the facility within that capacity envelope.

This interpretation is most relevant where the applicable service or interconnection arrangement recognizes an enforceable import limit supported by behind-the-meter resources and controllable demand. ICP-AI does not assume that all utilities or jurisdictions credit onsite resources against interconnection requirements in the same manner.

Future work can incorporate explicit network constraints, quantify the cost of enabling workload flexibility, investigate optimal BESS energy-to-power ratios through duration sensitivity analysis, account for PV and BESS degradation over the project lifetime, and test the framework using measured data center load profiles.

## 6. Conclusion

This paper developed ICP-AI, a developer-side planning framework that minimizes grid import capacity under a prescribed onsite investment budget while jointly sizing PV and BESS resources and incorporating aggregate workload flexibility. The results show that achievable interconnection capacity reduction depends strongly on the planning environment. At a $100 million budget, the reduction is about 6% under the conservative high load factor baseline, exceeds 10% when the same peak load profiles are paired with monthly average solar availability, and reaches about 13.3% for the more diurnal Mixed Business profile. Workload flexibility provides two distinct planning values: direct reduction of the grid capacity requirement and substitution for physical infrastructure. At a $10 million budget, the baseline case with 5% flexibility and a 1 h deferral window reduces BESS capacity from 15.30 to 4.87 MWh while increasing interconnection capacity reduction from 4.43% to 4.84%. The sensitivity results further show diminishing returns in direct capacity value, while storage substitution can remain substantial as flexibility increases. A full 8,760 h chronological analysis preserves the main interconnection capacity trends, supporting the robustness of the conclusions to temporal compression. Because the formulation depends on the facility load profile, onsite investment budget, and workload flexibility parameters rather than on a specific AI workload archetype, it can also be applied to other data center types with appropriately specified inputs. More generally, the same formulation may be extended to other large flexible loads with characterizable demand and shifting constraints. Overall, ICP-AI provides an investment-interconnection frontier for evaluating how onsite resources and workload flexibility can reduce data center grid capacity requirements and reshape the associated infrastructure portfolio.

## References

[1] International Energy Agency, Energy and AI, Paris, France, 2025.

[2] International Energy Agency, Electricity 2026, Paris, France, 2026.

[3] Federal Energy Regulatory Commission, 'FERC Launches Aggressive Targeted Action to Speed Large Load Integration,' Washington, DC, Jun. 18, 2026.

[4] U.S. Department of Energy, ‘Secretary Wright Acts to Unleash American Industry and Innovation with Newly Proposed Rules,’ Washington, DC, Oct. 23, 2025.

[5] P. Colangelo, A. K. Coskun, J. Megrue, et al., ‘AI data centres as grid-interactive assets,’ Nature Energy, vol. 11, pp. 254-261, 2026, doi: 10.1038/s41560-025-01927-1.

[6] C. Williams, P. Colangelo, A. Coskun, et al., ‘Power-Flexible AI Data Centers: A New Paradigm for Grid-Responsive Compute,’ arXiv:2606.25098, 2026.

[7] A. Caprara, Y. Yu, F. Teng, A. Junyent-Ferre, E. Bullich-Massague, and M. Aragues-Penalba, ‘Data center workload flexibility for power system demand response: Evidence from Alibaba traces,’ International Journal of Electrical Power & Energy Systems, vol. 178, Art. no. 111940, 2026.

[8] Y. Cao, M. Cheng, S. Zhang, H. Mao, P. Wang, C. Li, Y. Feng, and Z. Ding, ‘Data-driven flexibility assessment for internet data center towards periodic batch workloads,’ Applied Energy, vol. 324, Art. no. 119665, 2022.

[9] W. Zhang and V. M. Zavala, ‘Remunerating space-time, load-shifting flexibility from data centers in electricity markets,’ Applied Energy, vol. 326, Art. no. 119930, 2022.

[10] L. Liu, X. Shen, Z. Chen, Q. Sun, and R. Wennersten, ‘Optimal Energy Management of Data Center Micro-Grid Considering Computing Workloads Shift,’ IEEE Access, vol. 12, pp. 102061-102075, 2024.

[11] C. Guo, F. Luo, Z. Cai, Z. Y. Dong, and R. Zhang, ‘Integrated planning of internet data centers and battery energy storage systems in smart grids,’ Applied Energy, vol. 281, Art. no. 116093, 2021.

[12] Y. Zhang, H. Tang, H. Li, and S. Wang, ‘Unlocking the flexibilities of data centers for smart grid services: Optimal dispatch and design of energy storage systems under progressive loading,’ Energy, vol. 316, Art. no. 134511, 2025.

[13] I. Riepin, T. Brown, and V. M. Zavala, ‘Spatio-temporal load shifting for truly clean computing,’ Advances in Applied Energy, vol. 17, Art. no. 100202, 2025.

[14] Y. Zhang, B. Zou, X. Jin, Y. Luo, M. Song, Y. Ye, Q. Hu, Q. Chen, and A. C. Zambroni, ‘Mitigating power grid impact from proactive data center workload shifts: A coordinated scheduling strategy integrating synergistic traffic-data-power networks,’ Applied Energy, vol. 377, Art. no. 124697, 2025.

[15] H. Naoi, R. Delage, T. Nakata, and M. Kozai, ‘Integrating location planning and spatio-temporal workload shifting for 100% renewable energy data centers: A case study in Japan,’ Energy Strategy Reviews, vol. 61, Art. no. 101823, 2025.

[16] S. Liu, S. Shin, and D. Deka, ‘Watts vs. Bytes: Turning Data Centers into Grid Assets via Storage Compute Co-Optimization,’ arXiv:2605.16190, 2026.

[17] X. Lu, J. Qiu, J. Lin, S. An, M. Sun, and J. Zhao, ‘Battery-Assisted Operation of Hyperscale AI Data Centers under Connect-and-Manage Interconnection Practices,’ arXiv:2605.14105, 2026.

[18] Y. Mao, J. L. Mathieu, and V. Dvorkin, ‘Online Feedback Optimization of Energy Storage to Smooth Data Center Grid Impacts,’ arXiv:2603.20564, 2026.

[19] Y. Fan and J. Zhao, ‘Harnessing Flexible Spatial and Temporal Data Center Workloads for Grid Regulation Services,’ arXiv:2602.01508, 2026.

[20] J. Qu, Y. Yan, Q. Wang, and Z. Wang, ‘Flexibility-Enhanced Operation of Internet Data Centers with Operator-User Collaboration: A Bilevel Stackelberg Game Approach,’ IEEE Transactions on Industry Applications, early access, 2026.

[21] D. Kim, L. Dong, and L. Xie, ‘Flexibility-aware framework for efficient planner-initiated siting of data center,’ Nature Communications, vol. 17, Art. no. 6512, 2026.

[22] Y. Chen and X. Zheng, 'To Defer or To Shift? The Role of AI Data Center Flexibility on Grid Interconnection,' in Proc. 2026 ACM Sustainability Week, pp. 322-327, 2026.

[23] X. Lu and Q. Xu, ‘Grid Integration of Gigawatt-Scale AI Data Centers under Connect-and-Manage,’ arXiv:2605.14109, 2026.

[24] Q. Wang, T. Huang, Q. Liu, and J. Na, ‘Power-infrastructure expansion planning for training-oriented AI data centers under large-model scaling uncertainty,’ Energy and AI, vol. 25, Art. no. 100840, 2026.

[25] S. A. Birahim, ‘A net-grid-benefit test for interconnecting AI data centres,’ npj Environmental Social Sciences, vol. 1, Art. no. 8, 2026.

[26] M. Grahovac, S. J. Smith, A. Newkirk, E. Neill, Z. He, and V. Dementyeva, ‘Shape Maker: Data Center and Industrial Electrical Load Shape Generator,’ Lawrence Berkeley National Laboratory, software, 2026.

[27] A. P. Dobos, PVWatts Version 5 Manual, NREL/TP-6A20-62641, National Renewable Energy Laboratory, Golden, CO, 2014.

[28] National Renewable Energy Laboratory, 2024 Annual Technology Baseline: Electricity, Golden, CO, 2024.